\documentclass[11pt]{article}
\usepackage[a4paper,margin=27mm]{geometry}
\usepackage[T1]{fontenc}
\usepackage{lmodern}
\usepackage{amsmath,amssymb,amsthm,mathtools}
\usepackage{microtype}
\usepackage[colorlinks=true,linkcolor=blue,citecolor=blue,urlcolor=blue]{hyperref}
\newtheorem{theorem}{Theorem}[section]
\newtheorem{lemma}[theorem]{Lemma}
\newtheorem{proposition}[theorem]{Proposition}
\newtheorem{corollary}[theorem]{Corollary}
\theoremstyle{definition}
\newtheorem{definition}[theorem]{Definition}
\theoremstyle{remark}
\newtheorem{remark}[theorem]{Remark}
\newcommand{\F}{\mathbb F}
\newcommand{\C}{\mathcal C}
\newcommand{\A}{\mathcal A}
\newcommand{\wt}{\operatorname{wt}}
\newcommand{\supp}{\operatorname{supp}}
\newcommand{\ord}{\operatorname{ord}}
\newcommand{\PG}{\operatorname{PG}}
\newcommand{\GL}{\operatorname{GL}}
\newcommand{\ev}{\operatorname{ev}}
\newcommand{\CC}{\mathbb C}
\newcommand{\Wh}{\mathcal W}
\newcommand{\fall}[2]{(#1)_{\underline{#2}}}

\title{A Characterization of $q$-Ary $2$-Designs and a $3$-Design Obstruction for Doubly-Extended Reed-Solomon Codes}
\author{Guanghui~Zhang$^1$, Zhengbang Zha$^2$
\footnote{
E-mail addresses:
{\it zghui@squ.edu.cn (Guanghui Zhang); zhazhengbang@163.com (Zhengbang Zha).
}}
}

\date{\small
$1.$ School of Mathematics and Physics, Suqian University, Suqian, Jiangsu, 223800, China\\
$2.$ College of Mathematics and System Science, Xinjiang University, Urumqi 830046, China
}

\begin{document}
\maketitle
\begin{abstract}
Let $\mathcal{C}_k$ be the doubly extended Reed--Solomon code with
parameters $[q+1,k,q-k+2]_q$, where $q\ge3$ is a prime power and
$2\le k\le q$. We prove a conjecture of Wu and Zhou by showing that
the minimum-weight codewords of $\mathcal{C}_k$ do not form a
$q$-ary $3$-design when $2\le k\le q-1$. Using shortening, we extend
this obstruction to every nonempty weight class with
$3\le w<q+1$. We also establish that each individual nonempty
positive-weight class of $\mathcal{C}_k$ forms a $q$-ary $2$-design
if and only if $\gcd(k-1,q-1)=1$. Consequently, every nonempty
positive-weight class below full support has exact $q$-ary design
strength $2$ when this coprimality condition holds, and $1$ otherwise.
The three-design obstruction follows by applying a classical
subset-product character identity to a marginal covering count
for three prescribed nonzero symbols. The comparison between
weights rests on a covering-count formulation of the classical
MDS coset-weight relations. For every linear $[n,k,d]_q$ MDS code
with $2\le k\le n-1$, this formulation shows that the $q$-ary
$2$-design property at any one nonempty positive weight is
equivalent to the same property at every nonempty positive weight.
More generally, for $2\le t\le\min\{k,d\}$, we prove that the
$q$-ary $t$-design property at any $t-1$ distinct nonempty positive
weights forces it at all nonempty positive weights.
\end{abstract}

\section{Introduction}\label{sec:introduction}

The connection between error-correcting codes and combinatorial designs
has provided constructions and structural results in both subjects.
Incidence structures give rise to linear codes, while the supports of
codewords of a fixed weight may yield designs. The Assmus--Mattson
theorem \cite{AM} is a fundamental result in the latter direction.
Subsequent work has developed this relationship for many families of
codes; see \cite{HP} and the references therein.

The vector version of a design retains more information. In a $q$-ary
$t$-design, a block records both the positions and the values of its
nonzero coordinates. Every such design induces a classical $t$-design
of supports, with multiplicities, but the converse need not hold.
This notion originates in the work of Delsarte
\cite{DelsarteSchemes,DelsarteFour} and was developed by Assmus,
Goethals, and Mattson \cite{AGM}. More recently, Wu and Zhou
\cite{WZframework} established criteria for linear codes to hold
$q$-ary designs and applied automorphism groups to cases not covered
by their general criteria.

An important example is the doubly-extended Reed--Solomon (DRS) code
$\C_k$ with parameters $[q+1,k,q-k+2]_q$. Wu and Zhou showed that
its minimum-weight codewords form a $q$-ary $2$-design
\cite[Theorem 5.5]{WZframework} whenever
\[
 \gcd(k-1,q-1)=1.
\]
They subsequently proved necessity and established propagation from
minimum weight to every nonempty weight class of an MDS code
\cite{WZ26}. Thus the coprimality condition characterizes both the
minimum-weight property and the simultaneous property at all weights.

These results leave two further questions. First, can a higher-weight
class form a $q$-ary $2$-design when the minimum-weight class does not?
Second, can a DRS minimum-weight class have strength $3$?
The latter question was formulated as a nonexistence conjecture in
\cite[Remark 12]{WZ26}. Necessary divisibility conditions exclude many
parameters, but do not settle that conjecture for arbitrary prime powers.

In this paper, we answer both questions. First, we prove that, in every
linear MDS code with $2\le k\le n-1$, the $2$-design property at any
one nonempty weight is equivalent to the property at minimum weight.
The proof makes the entire pair-covering discrepancy explicit. This
gives the coprimality criterion at each individual weight of $\C_k$.
The same moment calculation also gives a higher-strength result:
for $2\le t\le\min\{k,d\}$, any $t-1$ distinct nonempty weight classes
that are $q$-ary $t$-designs force every nonempty weight class to be
a $q$-ary $t$-design.

Second, we prove that no minimum-weight class of $\C_k$ is a $q$-ary
$3$-design when $q\ge3$ and $2\le k\le q-1$. Fixing the symbols at
$0,1,\infty$ reduces a marginal covering count to products of subsets
of $\F_q^*\setminus\{1\}$. A faithful multiplicative character has
a nonzero coefficient for every relevant subset size, which contradicts
the required uniformity. Shortening then excludes $3$-designs at every
nonempty weight $3\le w<q+1$. Together with the individual-weight
$2$-design criterion, this determines the exact strength of every
nonempty weight class below full support.

The counting ingredients have established antecedents. MDS weight and
partition enumerators are treated in \cite{HP,ElKhamyMcEliece}, while
Bonneau's coset-weight formula \cite{Bonneau} and its reformulation in
\cite{DMP} explain the determination of higher coset weights from initial
ones. Our pair discrepancy identity is a covering-count specialization
of that framework when the shortened code has positive dimension.
Full-group subset counts are given by Li and Wan \cite{LiWan}; Kosters
\cite{Kosters} also treats deletion of the identity. We retain proofs
of the identities used here and formulate their consequences for
individual weight classes and for the DRS nonexistence conjecture.

The rest of the paper is organized as follows. Section~\ref{sec:prelim}
introduces the required notation and basic design and character facts.
Section~\ref{sec:moments} develops the MDS moment argument.
Section~\ref{sec:products} records the subset-product identities.
Section~\ref{sec:DRS} establishes the individual-weight classification
and exact covering discrepancies. Section~\ref{sec:min3} resolves the
minimum-weight conjecture, and Section~\ref{sec:shortening} extends the
obstruction by shortening. Section~\ref{sec:conclusion} concludes the paper.

\section{Preliminaries}\label{sec:prelim}

We introduce only the notation and basic facts needed throughout the
paper. The MDS interpolation lemma, the subset-product evaluations,
and the DRS automorphism argument are proved in the sections where
they are applied.

\subsection{Codes, weight classes, and \texorpdfstring{$q$}{q}-ary designs}

Let $q$ be a prime power, let $\F_q$ be the field of order $q$, and write
$\F_q^*=\F_q\setminus\{0\}$ and $[n]=\{1,\ldots,n\}$. For
$x=(x_1,\ldots,x_n)\in\F_q^n$, set
\[
 \supp(x)=\{i\in[n]:x_i\ne0\},\qquad
 \wt(x)=|\supp(x)|,\qquad d_H(x,y)=\wt(x-y).
\]
A linear $[n,k,d]_q$ code, with $1\le k\le n$, is a
$k$-dimensional subspace $\C$ of
$\F_q^n$ with $d=\min\{\wt(c):0\ne c\in\C\}$. It is
\emph{maximum distance separable} (MDS) if $d=n-k+1$.
The Singleton bound $d\le n-k+1$ follows by deleting any $d-1$
coordinates: two words with the same image would differ in at most
$d-1$ positions, so deletion is injective. Therefore
\[
 q^k=|\C|\le q^{n-d+1},\qquad k\le n-d+1.
\]
We use the notation
\[
 \A_w(\C)=\{c\in\C:\wt(c)=w\},\quad A_w=|\A_w(\C)|,
 \quad \Wh_t=\{x\in\F_q^n:\wt(x)=t\}.
\]
In particular, $|\Wh_t|=\binom nt(q-1)^t$.
The set $\A_w(\C)$ is called the \emph{weight-$w$ shell}
(or Hamming shell) of $\C$. It consists of all codewords
of Hamming weight $w$ and is nonempty precisely when
$A_w>0$.
Throughout, assertions about design strengths of codeword shells refer
to positive weights $d\le w\le n$; the zero word is excluded.

To distinguish the two design notions, recall that a classical
$t$-$(n,w,\lambda)$ design is a family of $w$-subsets of an $n$-set
in which each $t$-subset belongs to exactly $\lambda$ blocks.
The parameters $t$ and $\lambda$ are called the \textit{strength} and \textit{index}, repectively.
Repeated blocks are allowed when specified.

For the vector analogue,
write $c\succeq x$ if
\[
 c_i=x_i\qquad\text{for every }i\in\supp(x).
\]
Thus $c\succeq x$ means that $c$ covers $x$.

\begin{definition}
\label{def:design}
Let $\A\subseteq\Wh_w$ be nonempty and $1\le t\le w$. The set
$\A$ is a $q$-ary $t$-$(n,w,\lambda)$ design if
\[
 L_{\A}(x):=|\{c\in\A:c\succeq x\}|=\lambda
 \qquad\text{for every }x\in\Wh_t.
\]
For $\A=\A_w(\C)$, we write $L_w(x)$ and say that $\C$ holds
a $q$-ary $t$-design at weight $w$.
\end{definition}

The definition follows \cite{DelsarteSchemes,AGM,WZframework}.
Empty Hamming shells are not designs, and strengths greater than the block
weight are not considered. The \emph{exact strength} of a nonempty Hamming
shell is the largest $t$ for which it is a $q$-ary $t$-design.
If no positive strength exists, its exact strength is defined to be $0$.
If $\A$ is a $q$-ary $t$-design, its supports, counted once for each
vector of $\A$, form a classical design of index $(q-1)^t\lambda$:
for a fixed $t$-set, summing over its $(q-1)^t$ nonzero assignments
counts each containing block exactly once.

The following standard counting formulas will be used to compute
indices and to pass to smaller strengths. Their proofs are included
to fix the normalization.

\begin{lemma}
\label{lem:indices}
For every nonempty $\A\subseteq\Wh_w$ and $1\le t\le w$,
\begin{equation}\label{eq:index}
 \frac1{|\Wh_t|}\sum_{x\in\Wh_t}L_{\A}(x)
 =\frac{|\A|\binom wt}{(q-1)^t\binom nt}.
\end{equation}
If $\A$ is a $q$-ary $t$-design of index $\lambda_t$, then it is
a $q$-ary $i$-design for every $1\le i\le t$, with
\begin{equation}\label{eq:lowerindex}
 \lambda_i=\lambda_t(q-1)^{t-i}
                \frac{\binom{n-i}{t-i}}{\binom{w-i}{t-i}}.
\end{equation}
\end{lemma}
\begin{proof}
Count $\{(x,c)\in\Wh_t\times\A:c\succeq x\}$ in two ways.
Each $c$ covers exactly $\binom wt$ weight-$t$ vectors, since their
nonzero entries must equal those of $c$. Hence
\[
 \sum_{x\in\Wh_t}L_{\A}(x)=|\A|\binom wt,
\]
which gives \eqref{eq:index}.

Now fix $y\in\Wh_i$ and count pairs $(x,c)$ satisfying
$c\succeq x\succeq y$, $x\in\Wh_t$, $c\in\A$. There are
$\binom{n-i}{t-i}(q-1)^{t-i}$ possible $x$, each covered by
$\lambda_t$ blocks. Each $c\succeq y$ allows
$\binom{w-i}{t-i}$ such $x$. Therefore
\[
 L_{\A}(y)\binom{w-i}{t-i}
 =\lambda_t\binom{n-i}{t-i}(q-1)^{t-i}.
\]
The binomial factor on the left is positive, so $L_{\A}(y)$ is
constant and equals \eqref{eq:lowerindex}.
\end{proof}

For a Hamming shell, denote its average covering count in \eqref{eq:index} by
$\overline L_{w,t}$. This notation does not assert that the Hamming shell
is a design. If it is a design, its index is positive because
$A_w>0$ and $w\ge t$.

\subsection{Monomial maps and shortening}

The code operations used later preserve the relevant incidence relation.
For a permutation $\pi$ of $[n]$ and $v\in(\F_q^*)^n$, define
\[
 M_{\pi,v}(x)_i=v_i x_{\pi(i)}\qquad(i\in[n]).
\]
Such a map is \emph{monomial}; two codes are monomially equivalent
if one is the image of the other under such a map. A monomial
automorphism of $\C$ is a monomial map carrying $\C$ onto itself.
For any $c,x\in\F_q^n$,
\begin{equation}\label{eq:monomial}
 \wt(M_{\pi,v}x)=\wt(x),\qquad
 c\succeq x\ \Longleftrightarrow\ M_{\pi,v}c\succeq M_{\pi,v}x.
\end{equation}
Indeed, $v_i\ne0$ preserves zero coordinates and permits cancellation
in $v_i c_{\pi(i)}=v_i x_{\pi(i)}$. Consequently,
\[
 L_{M_{\pi,v}\A}(M_{\pi,v}x)=L_{\A}(x),
\]
so the $q$-ary design property is invariant under monomial equivalence.
This is the elementary action underlying the automorphism criterion
in \cite[Theorem 5.1]{WZframework}.

For $J\subseteq[n]$, let $\pi_{J^c}$ delete the coordinates in $J$.
The \emph{shortened code} is
\begin{equation}\label{eq:shortdefinition}
 \C_J=\{\pi_{J^c}(c):c\in\C,\ c_j=0\ (j\in J)\}.
\end{equation}
Deletion is injective on the displayed zero-coordinate subcode and
preserves its word weights. A coset of a linear code $B$ is a translate
$u+B=\{u+b:b\in B\}$.

For a map $\varphi:X\to Y$, the \emph{fibre over} $y\in Y$ is
\[
 \varphi^{-1}(y)=\{u\in X:\varphi(u)=y\}.
\]
In plain terms, it is the set of all inputs that give the same
specified output. In the code setting, consider the coordinate
restriction map
\[
 \rho_J:\C\longrightarrow\F_q^J,\qquad
 c\longmapsto(c_j)_{j\in J},
\]
where $\F_q^J$ denotes the set of tuples indexed by $J$.
For a prescribed tuple $a=(a_j)_{j\in J}$, its fibre is
\[
 \rho_J^{-1}(a)=\{c\in\C:c_j=a_j\text{ for every }j\in J\}.
\]
Thus a fibre consists of all codewords with fixed symbols at the
specified coordinates; the other coordinates may vary subject to
membership in $\C$. If this fibre is nonempty and $c_0$ belongs to it,
then $\rho_J^{-1}(a)=c_0+\ker\rho_J$ by linearity.

These definitions will relate sets of codewords with prescribed values
at specified coordinates to the classical coset-weight setting in
Section~\ref{sec:moments}; the design-shortening statement itself is
proved in Section~\ref{sec:shortening}.

\subsection{Cyclic characters and Fourier normalization}

Let $G=\langle\gamma\rangle$ be a cyclic group of order $m$.
A character is a homomorphism $\chi:G\to\CC^*$, and the character
group is denoted by $\widehat G$. Explicitly,
\[
 \chi_j(\gamma^a)=e^{2\pi i ja/m},\qquad 0\le j<m.
\]
The trivial character is $\chi_0=1$. For $m>1$, $\chi_1$ is
\emph{faithful}, meaning that its kernel is $\{1\}$; equivalently,
its order is $m$. The order of $\chi_j$ is $m/\gcd(j,m)$, so there
are $\varphi(h)$ characters of each order $h\mid m$, where
$\varphi$ is Euler's totient function. We apply this to the cyclic
group $\F_q^*$; see \cite[Chapters 2 and 5]{LN}.

For a function $H:G\to\CC$, use the transform
\[
 \widehat H(\chi)=\sum_{u\in G}H(u)\chi(u),\qquad
 \overline H=\frac1m\sum_{u\in G}H(u).
\]
The following standard identities underlie the character method of
\cite{Kosters}. We state their normalization explicitly because the
same coefficients later measure design discrepancies.

\begin{lemma}
\label{lem:Fourier}
For $u,v\in G$ and $\chi\in\widehat G$,
\begin{align}
 \sum_{u\in G}\chi(u)&=
   \begin{cases}m,&\chi=1,\\0,&\chi\ne1,\end{cases}\label{eq:charorth1}\\
 \sum_{\chi\in\widehat G}\chi(u)\overline{\chi(v)}
   &=m\,\mathbf1_{u=v}.\label{eq:charorth2}
\end{align}
Moreover,
\begin{align}
 H(u)&=\frac1m\sum_{\chi\in\widehat G}
              \widehat H(\chi)\overline{\chi(u)},\label{eq:inversionFourier}\\
 \sum_{u\in G}|H(u)-\overline H|^2
   &=\frac1m\sum_{\substack{\chi\in\widehat G\\\chi\ne1}}
                         |\widehat H(\chi)|^2.\label{eq:Parseval}
\end{align}
In particular, $H$ is constant if and only if
$\widehat H(\chi)=0$ for every nontrivial character.
\end{lemma}
\begin{proof}
Both orthogonality relations reduce to
\[
 \sum_{a=0}^{m-1}\zeta^a=
 \begin{cases}m,&\zeta=1,\\(1-\zeta^m)/(1-\zeta)=0,&\zeta\ne1,
 \end{cases}\qquad \zeta^m=1.
\]
Using \eqref{eq:charorth2}, the right side of
\eqref{eq:inversionFourier} becomes
\[
 \frac1m\sum_{v\in G}H(v)
       \sum_{\chi\in\widehat G}\chi(v)\overline{\chi(u)}=H(u).
\]
Similarly,
\begin{align*}
 \frac1m\sum_{\chi\in\widehat G}|\widehat H(\chi)|^2
 &=\frac1m\sum_{u,v\in G}H(u)\overline{H(v)}
          \sum_{\chi\in\widehat G}\chi(u)\overline{\chi(v)}\\
 &=\sum_{u\in G}|H(u)|^2.
\end{align*}
Since $\widehat H(1)=m\overline H$, subtracting
$m|\overline H|^2$ gives \eqref{eq:Parseval}. The final assertion
follows either from inversion or from this sum of nonnegative terms.
\end{proof}


For a polynomial or formal power series $P(T)$, the notation
$[T^r]P(T)$ denotes the coefficient of $T^r$ in $P(T)$.
More explicitly, if
\[
 P(T)=\sum_{j\ge0}a_jT^j,
\]
then
\[
 [T^r]P(T)=a_r\qquad(r\ge0).
\]
A term that is absent has coefficient zero. For example,
\[
 [T^2](1+3T+5T^2)=5,
 \qquad
 [T^3](1+3T+5T^2)=0.
\]
We also set $\binom ab=0$ when $a\ge0$ and $b\notin[0,a]$,
and interpret an empty sum as zero and an empty product as one.
These conventions apply also at the endpoints of the counting
formulas below.

\section{Zero-coordinate moments in an MDS code}\label{sec:moments}

Let $\C$ be a linear $[n,k,d]_q$ MDS code, so $d=n-k+1$.
We first derive the local counting identity that controls all subsequent
weight comparisons. The initial interpolation property is standard
\cite[Chapter 7]{HP}; the proof identifies exactly where the MDS
hypothesis enters.

\begin{lemma}
\label{lem:OA}
On any $s\le k$ distinct coordinate positions, each prescribed
$s$-tuple of field elements occurs in exactly $q^{k-s}$ codewords.
\end{lemma}
\begin{proof}
Let $G_0$ be a rank-$k$ generator matrix and let $I$ be a set of
$k$ coordinates. If the square submatrix $(G_0)_I$ were singular,
some $0\ne u\in\F_q^k$ would satisfy $u(G_0)_I=0$. But then
\[
 0\ne uG_0\in\C,\qquad \wt(uG_0)\le n-k=d-1,
\]
contrary to the minimum distance. For $|J|=s\le k$, extend $J$
to such an $I$. Thus $\operatorname{rank}(G_0)_J=s$, and the map
\[
 \F_q^k\longrightarrow\F_q^s,\qquad u\longmapsto u(G_0)_J
\]
is surjective. Each fibre, consisting of all message vectors $u$
that give the same prescribed coordinate tuple, is a coset of its
$(k-s)$-dimensional kernel and hence has size $q^{k-s}$.
\end{proof}

Fix $1\le t\le\min\{k,d\}$ and a weight-$t$ vector $x$. For
$0\le z\le k-1$, put
\[
 N_z(x)=L_{n-z}(x).
\]
Every codeword covering $x$ is nonzero, so it has at most $k-1$ zero
coordinates, all outside $\supp(x)$. Define its zero-coordinate moments by
\begin{align}
 F_j(x)
 &:=\sum_{\substack{J\subseteq[n]\setminus\supp(x)\\|J|=j}}
 |\{c\in\C:c\text{ covers }x,\ c_i=0\ (i\in J)\}|\notag\\
 &=\sum_{z=j}^{k-1}\binom zj N_z(x),
 \qquad 0\le j\le k-1.\label{eq:moments}
\end{align}
Lemma~\ref{lem:OA} gives
\begin{equation}\label{eq:lowmoments}
 F_j(x)=\binom{n-t}{j}q^{k-t-j},\qquad 0\le j\le k-t.
\end{equation}
For \eqref{eq:moments}, a word with $z$ zeros contributes once for
each of the $\binom zj$ subsets of its zero set. For
\eqref{eq:lowmoments}, the coordinates in $J$ and $\supp(x)$ are
disjoint, so the prescribed tuple occupies $j+t\le k$ positions;
Lemma~\ref{lem:OA} gives $q^{k-t-j}$ solutions for each of the
$\binom{n-t}{j}$ choices of $J$.
Thus all dependence on $x$ is confined to the last $t-1$ moments.
The next elementary inversion recovers exact zero counts from these
moments.

\begin{lemma}
\label{lem:inversion}
For every $0\le z\le k-1$,
\begin{equation}\label{eq:inversion}
 N_z(x)=\sum_{j=z}^{k-1}(-1)^{j-z}\binom jz F_j(x).
\end{equation}
\end{lemma}
\begin{proof}
Substitution of \eqref{eq:moments} makes the coefficient of $N_h(x)$
on the right equal to
\[
 \sum_{j=z}^{h}(-1)^{j-z}\binom jz\binom hj
 =\binom hz\sum_{j=z}^{h}(-1)^{j-z}\binom{h-z}{j-z}
 =\begin{cases}1,&h=z,\\0,&h>z.\end{cases}
\]
This proves the inversion formula.
\end{proof}

\subsection{An exact equivalence between individual weights}

For pair-covering counts, only one moment remains undetermined.
The resulting identity is a covering formulation of the classical
MDS coset enumerator relation discussed after the proof. Its nonzero
coefficient gives an equivalence between individual design weights.

\begin{theorem}
\label{thm:pair}
Let $\C$ be a linear $[n,k,d]_q$ MDS code with $2\le k\le n-1$.
For $d\le w\le n$, put $z=n-w$ and
\begin{equation}\label{eq:universal}
 U_{n,k}(z)=\sum_{j=z}^{k-2}(-1)^{j-z}\binom jz
                      \binom{n-2}{j}q^{k-j-2}.
\end{equation}
Then, for every weight-$2$ vector $x$,
\begin{equation}\label{eq:pairformula}
 L_w(x)=U_{n,k}(n-w)
       +(-1)^{w-d}\binom{k-1}{n-w}L_d(x).
\end{equation}
In particular, for any weight-$2$ vectors $x,y$,
\begin{equation}\label{eq:pairdifference}
 L_w(x)-L_w(y)
 =(-1)^{w-d}\binom{k-1}{n-w}\bigl(L_d(x)-L_d(y)\bigr).
\end{equation}
Consequently, the following are equivalent:\nopagebreak
\begin{enumerate}
\item[(1)] $\A_d(\C)$ is a $q$-ary $2$-design;
\item[(2)] $\A_w(\C)$ is a $q$-ary $2$-design for some $d\le w\le n$
      with $A_w>0$;
\item[(3)] $\A_w(\C)$ is a $q$-ary $2$-design for every $d\le w\le n$
      with $A_w>0$.
\end{enumerate}
The equivalence is valid separately for every choice of a nonempty
weight $w$ in the second assertion.
\end{theorem}
\begin{proof}
Set $t=2$ in \eqref{eq:lowmoments}. The moments $F_0,\ldots,F_{k-2}$
are independent of $x$, and
\[
 F_{k-1}(x)=N_{k-1}(x)=L_d(x).
\]
Equation~\eqref{eq:inversion} now gives \eqref{eq:pairformula}, since
$k-1-z=w-d$. Subtraction gives \eqref{eq:pairdifference}.
For $d\le w\le n$, we have $0\le n-w\le k-1$, and therefore
\[
 \binom{k-1}{n-w}>0.
\]
Constancy of $L_w$ is equivalent to constancy of $L_d$.
For a nonempty shell of weight at least $2$, its constant covering
number is positive by \eqref{eq:index}. This proves all assertions.
\end{proof}

\begin{remark}
For $k\ge3$, fix a weight-$2$ vector $x$ and put $J=\supp(x)$.
The fibre of codewords covering $x$ is precisely
\[
 \rho_J^{-1}(x|_J)
 =\{c\in\C:c_j=x_j\text{ for every }j\in J\}
 =\{c\in\C:c\succeq x\},
\]
where $x|_J=(x_j)_{j\in J}$. In other words, it consists of all
codewords agreeing with $x$ at its two nonzero positions.
Lemma~\ref{lem:OA} shows that this fibre is nonempty.
Puncturing this fibre at $J$ gives a coset $V_x$ of
the shortened $[N,k-2,d]_q$ MDS code $\C_J$, where $N=n-2$.
Indeed, the kernel of restriction to $J$ has dimension $k-2$;
shortening preserves weight, and the Singleton bound forces its
distance to equal $d$. Write
\[
 B_u(V_x)=|\{v\in V_x:\wt(v)=u\}|.
\]
Since both prescribed symbols are nonzero,
\[
 B_u(V_x)=0\quad(u<d-2),\qquad
 B_{w-2}(V_x)=L_w(x)\quad(d\le w\le n).
\]
Bonneau's formula \cite{Bonneau}, in the form
\cite[Theorem 1.1]{DMP}, therefore gives, for $u\ge d-1$,
\[
 B_u(V_x)=H_u(N,d,q)
       +(-1)^{u-d}\binom{N-d+2}{u-d+2}B_{d-2}(V_x),
\]
where
\[
 H_u(N,d,q)=A_u(\C_J)
          -(-1)^{u-d}\binom Nu\binom{u-1}{d-2}.
\]
The classical MDS weight distribution, proved in
Proposition~\ref{prop:weights}, shows that this term depends only
on $N,d,q,u$. With $u=w-2$, the coefficient is
\[
 (-1)^{w-d}\binom{k-1}{w-d}
 =(-1)^{w-d}\binom{k-1}{n-w}.
\]
Thus \eqref{eq:pairdifference} is a specialization of this classical
coset relation. The proof above also covers $k=2$ and derives the
constant term \eqref{eq:universal} directly. The design equivalence
uses the additional observation that the coefficient never vanishes
in the full range $d\le w\le n$.
\end{remark}

The theorem also quantifies the failure of the design property. If
$\overline L_{w,2}$ denotes the average in \eqref{eq:index}, averaging
\eqref{eq:pairformula} and subtracting gives
\begin{equation}\label{eq:centered}
 L_w(x)-\overline L_{w,2}
 =(-1)^{w-d}\binom{k-1}{n-w}
            \bigl(L_d(x)-\overline L_{d,2}\bigr).
\end{equation}
Thus the entire vector of discrepancies is preserved up to a nonzero
scalar, including its exact zeros and which entries have equal or opposite signs.
When $w-d$ is odd, all nonzero discrepancies reverse sign together.
If minimum-weight uniformity fails, every weight $d,\ldots,n$ must
occur: an empty shell would have identically zero covering counts and
would contradict \eqref{eq:pairdifference}.

\subsection{A higher-strength determination theorem}

The same inversion reveals a polynomial structure in the weight variable.
Write $\fall as=a(a-1)\cdots(a-s+1)$, with $\fall a0=1$.

\begin{theorem}
\label{thm:higher}
Let $\C$ be a linear MDS code and $2\le t\le\min\{k,d\}$.
For any weight-$t$ vectors $x,y$, there is a polynomial
$Q_{x,y}(Z)\in\mathbb Q[Z]$ of degree at most $t-2$ such that
\begin{equation}\label{eq:polynomial}
 L_{n-z}(x)-L_{n-z}(y)
 =(-1)^{k-1-z}\binom{k-1}{z}Q_{x,y}(z)
 \quad(0\le z\le k-1).
\end{equation}
If $\C$ holds $q$-ary $t$-designs at any $t-1$ distinct weights
in $\{d,\ldots,n\}$, it holds them at every nonempty positive weight.
\end{theorem}
\begin{proof}
Let $\Delta F_j=F_j(x)-F_j(y)$. Equation~\eqref{eq:lowmoments}
gives $\Delta F_j=0$ for $j\le k-t$. The remaining indices are
$j=k-1-s$, $0\le s\le t-2$. For $0\le z\le k-1$,
\[
 \frac{\binom{k-1-s}{z}}{\binom{k-1}{z}}
 =\frac{(k-1-s)!(k-1-z)!}{(k-1)!(k-1-s-z)!}
 =\frac{\fall{k-1-z}{s}}{\fall{k-1}{s}}
 \quad(z\le k-1-s).
\]
If $z>k-1-s$, both sides are zero. Hence \eqref{eq:inversion}
gives \eqref{eq:polynomial} with
\begin{equation}\label{eq:Q}
 Q_{x,y}(Z)=\sum_{s=0}^{t-2}(-1)^s
 \frac{\fall{k-1-Z}{s}}{\fall{k-1}{s}}\Delta F_{k-1-s}.
\end{equation}
Every denominator is nonzero: $0\le s\le t-2\le k-2$, so
$\fall{k-1}{s}$ is a product of positive integers (or is $1$).
The degree is at most $t-2$. The polynomial is over $\mathbb Q$,
not over $\F_q$; distinct integer weights remain distinct even when
they are congruent modulo the characteristic.
Suppose the $t-1$ design weights are $w_1,\ldots,w_{t-1}$ and put
$z_i=n-w_i$. Then $z_1,\ldots,z_{t-1}$ are distinct integers in
$[0,k-1]$. Equation~\eqref{eq:polynomial} and
$\binom{k-1}{z_i}>0$ imply $Q_{x,y}(z_i)=0$ for every $i$.
A nonzero polynomial of degree at most $t-2$ cannot have $t-1$
distinct roots, so $Q_{x,y}=0$. This holds for every pair $x,y$,
giving uniformity at all nonempty positive weights. The resulting
indices are positive by \eqref{eq:index}.
\end{proof}

At strength one, no undetermined moment remains. We record this
standard consequence, also stated in \cite[Theorem 4.2]{WZframework},
because it will give the lower bound in the exact-strength results.

\begin{corollary}
\label{cor:one}
Every nonempty positive-weight shell of a linear $[n,k,d]_q$ MDS
code is a $q$-ary $1$-design. For $z=n-w$, its index is
\[
 \lambda_{w,1}=\sum_{j=z}^{k-1}(-1)^{j-z}
                 \binom jz\binom{n-1}{j}q^{k-1-j}.
\]
In particular, $\lambda_{d,1}=\binom{n-1}{k-1}$.
\end{corollary}
\begin{proof}
Set $t=1$ in \eqref{eq:lowmoments}. Then
$F_j(x)=\binom{n-1}{j}q^{k-1-j}$ for every $0\le j\le k-1$,
independently of $x$. Substituting into \eqref{eq:inversion} gives
the formula and proves uniformity. For $w=d$, we have $z=k-1$,
so the sum has only its last term.
\end{proof}

\section{Character coefficients for subset products}\label{sec:products}

We next record the classical counting ingredients in a form suited to
the code. Let $G$ be a cyclic group of order $m$, written multiplicatively.
For $D\subseteq G$, define
\[
 N_D(r,u)=\left|\left\{S\subseteq D:|S|=r,
                            \prod_{a\in S}a=u\right\}\right|.
\]
The empty product is the identity. If $\chi$ is a complex character of
$G$, its transform is
\begin{equation}\label{eq:fourier}
 E_{D,r}(\chi)=\sum_{u\in G}N_D(r,u)\chi(u)
 =[T^r]\prod_{a\in D}(1+\chi(a)T).
\end{equation}
The function $N_D(r,\cdot)$ is constant if and only if all its
nontrivial character coefficients vanish, by Lemma~\ref{lem:Fourier}.
The coefficient identity follows by expanding
\[
 \prod_{a\in D}(1+\chi(a)T)
 =\sum_{S\subseteq D}\chi\left(\prod_{a\in S}a\right)T^{|S|}.
\]
We first evaluate this polynomial for the full group.

\begin{lemma}
\label{lem:fullgroup}
If $\ord(\chi)=h\mid m$ and $0\le r\le m$, then
\begin{equation}\label{eq:fullcoefficient}
 E_{G,r}(\chi)=
 \begin{cases}
 (-1)^{r+r/h}\binom{m/h}{r/h},&h\mid r,\\
 0,&h\nmid r.
 \end{cases}
\end{equation}
For $1\le r\le m$, $N_G(r,u)$ is independent of $u$ if and only
if $\gcd(r,m)=1$. In that case its value is $\binom mr/m$.
\end{lemma}
\begin{proof}
Every $h$th root of unity occurs $m/h$ times among the values of $\chi$.
Consequently,
\begin{equation}\label{eq:product}
 \prod_{a\in G}(1+\chi(a)T)
 =\left(\prod_{\zeta^h=1}(1+\zeta T)\right)^{m/h}
 =\bigl(1-(-T)^h\bigr)^{m/h}.
\end{equation}
The binomial expansion is
\[
 \bigl(1-(-T)^h\bigr)^{m/h}
 =\sum_{j=0}^{m/h}(-1)^{j+hj}\binom{m/h}{j}T^{hj}.
\]
The coefficient of $T^r$ vanishes unless $h\mid r$; otherwise
$j=r/h$, giving the sign $(-1)^{r+r/h}$ in
\eqref{eq:fullcoefficient}.
If $\gcd(r,m)=1$, no nontrivial character order divides $r$.
Conversely, if $g=\gcd(r,m)>1$, a character of order $g$ has a
nonzero coefficient. Fourier inversion proves the criterion and the
average gives the stated value. When $m=1$ the conclusion is immediate.
\end{proof}

This identity is part of the character approach in \cite{Kosters} and
the cyclic specialization of \cite{LiWan}; the uniformity criterion
also appears as Lemma 7 of \cite{WZ26}.
For example, if $G=\langle\gamma\rangle$ and $u=\gamma^a$,
Fourier inversion, grouping characters by their order, yields the
classical Ramanujan-sum formula \cite[Corollary 1.2]{LiWan}
\begin{equation}\label{eq:ramanujan}
 N_G(r,\gamma^a)=\frac1m
 \sum_{h\mid\gcd(m,r)}(-1)^{r+r/h}\binom{m/h}{r/h}c_h(a),
 \qquad
 c_h(a)=\sum_{\substack{1\le b\le h\\\gcd(b,h)=1}}
          e^{2\pi i ab/h}.
\end{equation}
Indeed, the characters of order $h$ take $\gamma$ to the primitive
$h$th roots of unity. Their contribution in the inverse transform is
$c_h(-a)=c_h(a)$, since $b\mapsto h-b$ permutes the reduced residue
classes. We will need the corresponding deleted-identity coefficient.

\begin{lemma}
\label{lem:deleted}
Assume $m\ge2$, put $D=G\setminus\{1\}$, and let
$0\le r\le m-1$. For a character $\chi$ of order $h>1$,
\begin{equation}\label{eq:deletedcoefficient}
 E_{D,r}(\chi)
 =(-1)^{r+\lfloor r/h\rfloor}
       \binom{m/h-1}{\lfloor r/h\rfloor}\ne0.
\end{equation}
In particular, for a faithful character, of order $m$,
\begin{equation}\label{eq:faithful}
 E_{D,r}(\chi)=(-1)^r.
\end{equation}
Therefore $N_D(r,u)$ is never constant in $u$ in this range.
\end{lemma}
\begin{proof}
The character is defined on $G$, not on $D$, which is not a group.
Removing the identity, whose character value is $1$, divides
\eqref{eq:product} by $1+T$:
\[
 \prod_{a\in D}(1+\chi(a)T)
 =\frac{(1-(-T)^h)^{m/h}}{1+T}.
\]
Put $M=m/h$ and $v=\lfloor r/h\rfloor$. Coefficient extraction gives
\[
 E_{D,r}(\chi)
 =(-1)^r\sum_{j=0}^{v}(-1)^j\binom Mj
 =(-1)^{r+v}\binom{M-1}{v}.
\]
For clarity, Pascal's identity gives the cancellation
\begin{align*}
 \sum_{j=0}^{v}(-1)^j\binom Mj
 &=\sum_{j=0}^{v}(-1)^j\binom{M-1}{j}
   +\sum_{j=1}^{v}(-1)^j\binom{M-1}{j-1}\\
 &=(-1)^v\binom{M-1}{v}.
\end{align*}
Since $r\le m-1$, we have $0\le v\le M-1$,
so the binomial coefficient is positive. For $h=m$, $v=0$.
Nonconstancy follows from character orthogonality. The coefficient
formula is the cyclic Fourier form of \cite[Theorem 1.3]{Kosters}.
The same result is numbered Theorem 3.1 in arXiv:1112.6294v1;
the published numbering is also reproduced in
\cite[Chapter 5, Theorem 1.3]{KostersThesis}.
\end{proof}

\section{The individual-weight classification for DRS codes}\label{sec:DRS}

We now apply the two preceding sections to a concrete evaluation code.
Let $2\le k\le q$, put $r=k-1$, and define
\begin{equation}\label{eq:DRS}
 \C_k=\{(f(a))_{a\in\F_q}\mathbin\Vert f(\infty):
                         f\in\F_q[X],\ \deg f\le r\},
 \qquad f(\infty)=[X^r]f.
\end{equation}
Here $\Vert$ appends the last coordinate. In this paper,
\emph{doubly-extended Reed-Solomon code} (or \emph{projective Reed--Solomon code}) refers to this same evaluation code; the terminology
does not denote two different code families. This is the DRS code in
\cite{WZ26}; its coordinates are indexed by
$\PG(1,q)=\F_q\cup\{\infty\}$. We include the elementary
parameter argument. If all finite evaluations vanish, then a polynomial
of degree at most $r\le q-1$ has $q$ roots and is zero. Thus the
evaluation map is injective and $\dim\C_k=r+1=k$. For $f\ne0$,
the number of zero coordinates satisfies
\[
 |\{a\in\F_q:f(a)=0\}|+\mathbf1_{f(\infty)=0}
 \le
 \begin{cases}
 r,&\deg f=r,\\
 (r-1)+1=r,&\deg f<r.
 \end{cases}
\]
Consequently $\wt(c_f)\ge q+1-r=q-k+2$. Choosing $r$ distinct
finite roots gives equality, so $\C_k$ is an $[q+1,k,q-k+2]_q$
MDS code. It remains to express its local minimum-weight counts in
the subset-product notation.

\begin{lemma}
\label{lem:projective}
The monomial automorphism group of $\C_k$ induces a transitive action
on ordered pairs of distinct coordinate positions. At the pair
$(0,\infty)$, the minimum-weight covering count for prescribed
nonzero values $(a,b)$ is
\begin{equation}\label{eq:rootcount}
 L_d^{0,\infty}(a,b)
 =N_{\F_q^*}\bigl(r,(-1)^r a/b\bigr).
\end{equation}
\end{lemma}
\begin{proof}
Homogenize $f$ to the binary form $F(X,Y)=Y^r f(X/Y)$ of degree $r$.
Set $p_a=(a,1)^{\mathsf T}$ for $a\in\F_q$ and
$p_\infty=(1,0)^{\mathsf T}$. Then
$\ev(F)=(F(p_P))_{P\in\PG(1,q)}$ is exactly \eqref{eq:DRS}.
Let $\GL(2,q)$ denote the group of invertible $2\times2$ matrices
over $\F_q$. For $A\in\GL(2,q)$, write uniquely
\[
 A p_P=\rho_A(P)p_{\sigma_A(P)},\qquad
 \rho_A(P)\in\F_q^*,\quad \sigma_A(P)\in\PG(1,q).
\]
The map $F\mapsto F\circ A$ is invertible, with inverse
$F\mapsto F\circ A^{-1}$. Homogeneity gives
\[
 \ev(F\circ A)_P
 =F(Ap_P)=\rho_A(P)^r F(p_{\sigma_A(P)}).
\]
Hence this substitution induces a monomial automorphism of $\C_k$.
For any ordered pairs of distinct points $(P,Q)$ and $(P',Q')$,
both $(p_P,p_Q)$ and $(p_{P'},p_{Q'})$ are bases of $\F_q^2$.
The invertible matrix sending the first basis to the second induces
$P\mapsto P'$ and $Q\mapsto Q'$. This proves the required
transitivity. Equation~\eqref{eq:monomial} then transfers covering
counts between pairs; its nonzero coordinate factors biject the
possible symbol assignments.

Now prescribe $f(\infty)=b\ne0$. Then $\deg f=r$. For weight $d$,
$f$ must have exactly $r$ distinct finite roots. Prescribing
$f(0)=a\ne0$ excludes zero from its root set, so
\[
 f(X)=b\prod_{u\in S}(X-u),\qquad
 S\subseteq\F_q^*,\quad |S|=r,
 \quad \prod_{u\in S}u=(-1)^r a/b.
\]
This correspondence is bijective and proves \eqref{eq:rootcount}.
\end{proof}

\begin{theorem}
\label{thm:DRS2}
Let $q$ be a prime power and $2\le k\le q$. For every
$d\le w\le q+1$ such that $\A_w(\C_k)\ne\varnothing$,
\begin{equation}\label{eq:classification}
 \A_w(\C_k)\text{ is a $q$-ary $2$-design}
 \quad\Longleftrightarrow\quad \gcd(k-1,q-1)=1.
\end{equation}
If the coprimality condition fails, no nonempty weight class is a
$q$-ary $2$-design, and every weight from $d$ to $q+1$ occurs.
\end{theorem}
\begin{proof}
Lemma~\ref{lem:fullgroup} and \eqref{eq:rootcount} show that uniformity
at $(0,\infty)$ is equivalent to the coprimality condition. Under
that condition the common value is $\binom{q-1}{k-1}/(q-1)$.
Lemma~\ref{lem:projective} transfers it to every coordinate pair.
Thus the minimum-weight criterion holds, recovering
\cite[Theorem 8]{WZ26}. Theorem~\ref{thm:pair} proves its equivalence
with the property at each individual nonempty weight and gives the
assertion about occurring weights when uniformity fails.
\end{proof}

The distinction between an all-weight characterization and
\eqref{eq:classification} is substantive: the latter rules out an
isolated higher-weight $2$-design whenever the coprimality condition
fails. It also supplies its exact obstruction. Put $m=q-1$, $r=k-1$,
and $z=q+1-w$. Combining \eqref{eq:pairformula} and
\eqref{eq:rootcount} gives
\begin{equation}\label{eq:DRSlocal}
 L_w^{0,\infty}(a,b)
 =U_{q+1,k}(z)+(-1)^{w-d}\binom rz
                    N_G\bigl(r,(-1)^r a/b\bigr).
\end{equation}
The character coefficients therefore measure the obstruction at each
weight, rather than only deciding whether it vanishes.

\begin{proposition}
\label{prop:variance}
Put $m=q-1$, $r=k-1$, $G=\F_q^*$, and $z=q+1-w$ for
$d\le w\le q+1$. Define
\[
 \mu_w=\frac1m\sum_{u\in G}L_w^{0,\infty}((-1)^r u,1).
\]
Then
\begin{equation}\label{eq:variance}
 \sum_{u\in G}\left|L_w^{0,\infty}((-1)^r u,1)-\mu_w\right|^2
 =\frac{\binom rz^2}{m}
   \sum_{\substack{h\mid\gcd(r,m)\\h>1}}
           \varphi(h)\binom{m/h}{r/h}^{\!2}.
\end{equation}
This quantity vanishes if and only if $\gcd(r,m)=1$.
\end{proposition}
\begin{proof}
The subsets counted by $N_G(r,u)$ partition all $r$-subsets of $G$,
so their mean is $\nu=\binom mr/m$. Equation~\eqref{eq:DRSlocal}
therefore gives
\begin{align*}
 \mu_w&=U_{q+1,k}(z)+(-1)^{w-d}\binom rz\nu,\\
 L_w^{0,\infty}((-1)^r u,1)-\mu_w
 &=(-1)^{w-d}\binom rz\bigl(N_G(r,u)-\nu\bigr).
\end{align*}
Squaring and summing, then applying \eqref{eq:Parseval}, yields
\[
 \sum_{u\in G}\left|L_w^{0,\infty}((-1)^r u,1)-\mu_w\right|^2
 =\frac{\binom rz^2}{m}
    \sum_{\substack{\chi\in\widehat G\\\chi\ne1}}
                    |E_{G,r}(\chi)|^2.
\]
For each $h\mid m$, there are $\varphi(h)$ characters of order $h$.
Their coefficients vanish unless $h\mid r$; in the remaining cases
their squared absolute values are $\binom{m/h}{r/h}^2$ by
Lemma~\ref{lem:fullgroup}. This proves \eqref{eq:variance}.
All terms are nonnegative, and every term indexed by a divisor
$h>1$ of $\gcd(r,m)$ is positive. Since $\binom rz>0$, the final
criterion follows.
\end{proof}

When the discrepancy vanishes, the design index is determined by
the weight distribution. For completeness, we derive the classical
MDS distribution \cite[Chapter 7]{HP}; the same fixed-support count
also follows from \cite[Corollary 5]{ElKhamyMcEliece}.

\begin{proposition}
\label{prop:weights}
Let $\C$ be a linear $[n,k,d]_q$ MDS code. For $d\le w\le n$,
\begin{equation}\label{eq:MDSweights}
 A_w=(q-1)\binom nw
            \sum_{j=0}^{w-d}(-1)^j\binom{w-1}{j}q^{w-d-j}.
\end{equation}
In particular, $A_d=(q-1)\binom nd$.
\end{proposition}
\begin{proof}
Fix a set $S\subseteq[n]$ of size $w$, and let $M(S)$ count words
whose support is exactly $S$. A word with support contained in
$S\setminus J$, where $J\subseteq S$ and $|J|=j$, vanishes on
$n-w+j$ coordinates. If $n-w+j\le k$, Lemma~\ref{lem:OA}
counts $q^{k-(n-w+j)}=q^{w-d+1-j}$ such words. If
$n-w+j\ge k$, only the zero word is possible. Inclusion--exclusion
therefore gives
\begin{align*}
 M(S)
 &=\sum_{j=0}^{w}(-1)^j\binom wj q^{\max\{0,w-d+1-j\}}\\
 &=\sum_{j=0}^{w-d}(-1)^j\binom wj
                         \bigl(q^{w-d+1-j}-1\bigr).
\end{align*}
For the second equality, use $\sum_{j=0}^{w}(-1)^j\binom wj=0$;
the terms with $j\ge w-d+1$ contribute no additional power above
$q^0=1$. Put $\ell=w-d$ and
\[
 R=\sum_{j=0}^{\ell}(-1)^j\binom{w-1}{j}q^{\ell-j}.
\]
Pascal's identity and the alternating-binomial cancellation give
\begin{align*}
 \sum_{j=0}^{\ell}(-1)^j\binom wj q^{\ell+1-j}
   &=(q-1)R+(-1)^\ell\binom{w-1}{\ell},\\
 \sum_{j=0}^{\ell}(-1)^j\binom wj
   &=(-1)^\ell\binom{w-1}{\ell}.
\end{align*}
Subtracting gives $M(S)=(q-1)R$, independently of $S$.
There are $\binom nw$ choices of $S$, proving \eqref{eq:MDSweights}.
For $w=d$, the sum has the single term $1$.
\end{proof}

Substituting this classical distribution into the general index
formula recovers the DRS index already obtained by Wu and Zhou.

\begin{corollary}[Corollary 10; Wu-Zhou \cite{WZ26}]
If $\gcd(k-1,q-1)=1$ and $\A_w(\C_k)\ne\varnothing$ for
$d\le w\le q+1$, its $q$-ary $2$-design index is
\begin{equation}\label{eq:weightindex}
 \lambda_w=\frac1{q-1}\binom{q-1}{w-2}
     \sum_{j=0}^{w-d}(-1)^j\binom{w-1}{j}q^{w-d-j}.
\end{equation}
\end{corollary}
\begin{proof}
Theorem~\ref{thm:DRS2} supplies the design property. With $n=q+1$,
\eqref{eq:index} and \eqref{eq:MDSweights} give
\begin{align*}
 \lambda_w
 &=\frac{A_w\binom w2}{(q-1)^2\binom{q+1}2}\\
 &=\frac1{q-1}\binom{q-1}{w-2}
       \sum_{j=0}^{w-d}(-1)^j\binom{w-1}{j}q^{w-d-j},
\end{align*}
where the last equality uses
$\binom{q+1}{w}\binom w2=\binom{q+1}2\binom{q-1}{w-2}$.
\end{proof}

\section{Resolution of the minimum-weight three-design conjecture}\label{sec:min3}

We now turn to three prescribed nonzero symbols. The extra coordinate
removes the identity from the root-product domain, and
Lemma~\ref{lem:deleted} makes uniformity impossible.

\begin{theorem}
\label{thm:no3min}
For every prime power $q\ge3$ and every $2\le k\le q-1$,
the minimum-weight codewords of the DRS code $\C_k$ do not form
a $q$-ary $3$-design. This proves the conjecture in
\cite[Remark 12]{WZ26}.
\end{theorem}
\begin{proof}
Put $r=k-1$, so $1\le r\le q-2$, and put
$D=\F_q\setminus\{0,1\}=\F_q^*\setminus\{1\}$.
For $a,b\in\F_q^*$, let
\[
 T(a,b)=|\{c_f\in\A_d(\C_k): f(0)=a,\ f(1)=b,
                                         \ f(\infty)=1\}|.
\]
If $\A_d(\C_k)$ were a $q$-ary $3$-design, $T(a,b)$ would be a
constant $\lambda$. A word counted by $T(a,b)$ is uniquely represented by
\[
 f(X)=\prod_{u\in S}(X-u),\qquad S\subseteq D,\quad |S|=r,
\]
with
\[
 a=(-1)^r\prod_{u\in S}u,\qquad b=\prod_{u\in S}(1-u).
\]
The second product is automatically nonzero. Summing over all $b\ne0$
therefore gives
\begin{equation}\label{eq:triplemarginal}
 \sum_{b\in\F_q^*}T(a,b)=N_D\bigl(r,(-1)^r a\bigr).
\end{equation}
Choose a faithful multiplicative character $\chi$ of $\F_q^*$.
It is nontrivial because $q-1\ge2$. The character is complex-valued, even when $q$ has characteristic $2$.
Thus the integer $(-1)^r$ below is nonzero in $\CC$.
The substitution $a\mapsto(-1)^r a$ permutes $\F_q^*$, so
\eqref{eq:charorth1}
applies to the resulting weighted sum.
Uniformity would imply
\[
 \sum_{a,b\in\F_q^*}T(a,b)\chi((-1)^r a)
 =(q-1)\lambda\sum_{a\in\F_q^*}\chi((-1)^r a)=0.
\]
On the other hand, \eqref{eq:triplemarginal} and
\eqref{eq:faithful} give
\[
 \sum_{a,b\in\F_q^*}T(a,b)\chi((-1)^r a)
 =\sum_{\substack{S\subseteq D\\|S|=r}}
                 \chi\left(\prod_{u\in S}u\right)
 =(-1)^r\ne0.
\]
This contradiction proves the theorem. It requires neither a
coprimality assumption nor an integrality test.
\end{proof}

The proof has a useful quantitative form. For any character of order
$h>1$, the last nonzero certificate is
\begin{equation}\label{eq:generalcertificate}
 (-1)^{r+\lfloor r/h\rfloor}
       \binom{(q-1)/h-1}{\lfloor r/h\rfloor}.
\end{equation}
Thus every nontrivial multiplicative character detects a failure of
uniformity in the marginal \eqref{eq:triplemarginal}.
For the parameters $q=1849$, $k=828$ or $k=1022$ highlighted in
\cite[Remark 12]{WZ26}, the faithful-character certificate is $-1$ in
both cases. These parameters are excluded even though the stated
integrality and coprimality tests are satisfied.

Combining this obstruction with the pair criterion now determines
the exact strength at minimum weight.

\begin{corollary}
\label{cor:minstrength}
For $q\ge3$ and $2\le k\le q$, the maximum strength of the
nonempty minimum-weight class of $\C_k$ is
\[
 \begin{cases}
 2,&\gcd(k-1,q-1)=1,\\
 1,&\gcd(k-1,q-1)>1.
 \end{cases}
\]
\end{corollary}
\begin{proof}
Corollary~\ref{cor:one} gives strength at least $1$, and
Theorem~\ref{thm:DRS2} settles strength $2$. For $k\le q-1$,
Theorem~\ref{thm:no3min} excludes strength $3$ and hence, by
Lemma~\ref{lem:indices}, every larger strength. For $k=q$,
$d=2$ and $\gcd(k-1,q-1)=q-1>1$, so the strength is $1$.
\end{proof}

\section{Extension to all weights below full support}\label{sec:shortening}

The minimum-weight obstruction can be propagated after shortening.
The next lemma is the strength-$3$ case of the shortening property
in \cite[Lemma 3.7]{WZframework}, where the result is attributed to
\cite{AGM}. We include a direct proof, including strict positivity
of the shortened index. This positivity will be essential
when we apply the pair-covering rigidity theorem to a shortened DRS code.

\begin{lemma}
\label{lem:shortendesign}
Suppose $\A_w(\C)$ is a $q$-ary $3$-design of length $n$ with
$3\le w<n$. Fix a coordinate $\ell$ and shorten $\C$ at $\ell$,
obtaining $\C_{\{\ell\}}$. Then
$\A_w(\C_{\{\ell\}})$ is a nonempty $q$-ary $2$-design.
\end{lemma}
\begin{proof}
Let $\lambda_3>0$ be the triple index and $\lambda_2$ the induced
pair index. Fix a weight-$2$ vector $x$ supported away from $\ell$.
Among the $\lambda_2$ covering words, those nonzero at $\ell$
partition into $q-1$ sets according to their value $a\in\F_q^*$.
Each set covers the weight-$3$ vector $x+a e_\ell$, so has size
$\lambda_3$, where $e_\ell$ is the vector with $1$ in coordinate
$\ell$ and zeros elsewhere. The number with zero at $\ell$ is thus
\begin{equation}\label{eq:shortindex}
 \lambda_2-(q-1)\lambda_3
 =(q-1)\lambda_3\frac{n-w}{w-2}>0,
\end{equation}
because \eqref{eq:lowerindex} gives
$\lambda_2=(q-1)(n-2)\lambda_3/(w-2)$.
The count is independent of the prescribed pair. Deleting the zero
coordinate preserves weight and covering, proving the assertion.
\end{proof}

We next examine the shortened code itself. Its additional forced root
replaces a full-group subset count by a deleted-identity count, and
therefore destroys the $2$-design property already at minimum weight.

\begin{proposition}
\label{prop:shortened}
Let $q\ge3$ and $3\le k\le q$. Shorten $\C_k$ at the coordinate
$1\in\F_q$, and call the resulting code $\mathcal S_k$.
Then $\mathcal S_k$ is an $[q,k-1,q-k+2]_q$ MDS code.
Every weight $d,\ldots,q$ occurs, and no such weight class is a
$q$-ary $2$-design.
\end{proposition}
\begin{proof}
The map $f\mapsto f(1)$ on polynomials of degree at most $k-1$
is a nonzero linear functional, since it takes the constant polynomial
$1$ to $1$. Its kernel has dimension $k-1$. Evaluation is injective,
and deleting a coordinate already equal to zero is also injective;
hence $\dim\mathcal S_k=k-1$. All its nonzero words have weight
at least $d$, since deletion preserves their original weights.
Choose $k-1$ distinct finite roots including $1$ and take their monic
product. This is possible because $k-1\le q-1$ and produces a
word of weight $d$ that is zero at $1$. Thus
\[
 d(\mathcal S_k)=d=q-k+2=q-(k-1)+1,
\]
proving the asserted MDS parameters.

Put $r=k-1$. Prescribe nonzero values $f(0)=a$ and $f(\infty)=1$.
Its minimum-weight words are precisely
\[
 f(X)=(X-1)\prod_{u\in S}(X-u),\qquad
 S\subseteq\F_q^*\setminus\{1\},\quad |S|=r-1.
\]
Their count is
\[
 N_{\F_q^*\setminus\{1\}}\bigl(r-1,(-1)^r a\bigr).
\]
Here $1\le r-1=k-2\le q-2$, and the ambient cyclic group has
order $q-1\ge2$. Lemma~\ref{lem:deleted} applies; in particular,
for a faithful character $\chi$,
\[
 \sum_{a\in\F_q^*}
 N_{\F_q^*\setminus\{1\}}(r-1,(-1)^r a)\,
          \chi((-1)^r a)=(-1)^{r-1}\ne0.
\]
If the count were independent of $a$, orthogonality would make this
sum zero. Thus the count is not constant in $a$. Notice that $\mathcal S_k$
is not the full-length DRS code $\C_{k-1}$: its length is $q$,
so Theorem~\ref{thm:DRS2} is not being applied to it.
The shortened code therefore fails the minimum-weight $2$-design
condition. Since $2\le k-1\le q-1$ and $d\ge2$,
Theorem~\ref{thm:pair} applies to
$\mathcal S_k$ and proves both remaining assertions.
\end{proof}

The two preceding results are incompatible with a $3$-design of
weight smaller than the length in the original code. This yields the
following extension of Theorem~\ref{thm:no3min}.

\begin{theorem}
\label{thm:no3weights}
Let $q\ge3$ and $2\le k\le q$. If $\A_w(\C_k)\ne\varnothing$
and $3\le w\le q$, then $\A_w(\C_k)$ is not a $q$-ary $3$-design.
\end{theorem}
\begin{proof}
If $k\ge3$, a $3$-design at such a weight would, by
Lemma~\ref{lem:shortendesign}, give a $2$-design of the same weight
in $\mathcal S_k$, contradicting Proposition~\ref{prop:shortened}.
If $k=2$, every nonzero codeword has weight $q$: a nonconstant
linear polynomial has one finite zero and a nonzero infinity value,
whereas a nonzero constant has no finite zeros and is zero at infinity.
Thus only the minimum-weight class occurs, and
Theorem~\ref{thm:no3min} applies.
\end{proof}

Together with the individual-weight classification, this obstruction
leaves only strengths one and two.

\begin{corollary}
\label{cor:strength}
Let $q\ge3$, $2\le k\le q$, and $d\le w\le q$ with
$\A_w(\C_k)\ne\varnothing$. Its maximum $q$-ary design strength is
$2$ when $\gcd(k-1,q-1)=1$, and $1$ otherwise.
\end{corollary}
\begin{proof}
Combine Corollary~\ref{cor:one} with Theorems~\ref{thm:DRS2}
and~\ref{thm:no3weights}, using Lemma~\ref{lem:indices} to exclude
larger strengths. If $w=2$, then $d\le2$ forces $k=q$ and the
coprimality condition fails, so that case also has strength $1$.
\end{proof}

\begin{remark}
\label{rem:boundary}
Theorem~\ref{thm:DRS2} includes $w=q+1$ whenever that shell is
nonempty. Theorem~\ref{thm:no3weights} does not include $w=q+1$:
the positive factor $n-w$ in \eqref{eq:shortindex} becomes zero.
Thus this paper does not claim a classification of $3$-designs in
the full-support shell. If the coprimality condition fails, that shell
is already excluded at strength $2$. In the coprime case, its possible
strength $3$ requires a separate argument. The minimum-weight
conjecture is nevertheless completely resolved by
Theorem~\ref{thm:no3min}.
\end{remark}

\begin{remark}
\label{rem:endpoints}
The nonemptiness qualifications are necessary. For $k=2$, every
nonzero word has weight $q$, so all higher shells are empty.
For $q=2$, the only dimension in the classification range is $k=2$;
the code is
\[
 \C_2=\{000,011,101,110\},
\]
and its weight-$2$ shell is a binary $2$-design of exact strength $2$.
The three-design statements assume $q\ge3$ and do not apply to this
weight-$2$ shell. For $q\ge3$ and $k=q$, the minimum distance is $2$
and $\gcd(k-1,q-1)=q-1>1$, so the individual-weight criterion excludes
strength $2$ at every nonempty shell. None of these boundary cases
requires treating an empty shell as a design or a strength greater
than the block weight.
\end{remark}

Finally, \eqref{eq:monomial} shows that the DRS conclusions also
hold for every code monomially equivalent to $\C_k$, including all
its nonzero column-scaled versions.

\section{Conclusion}\label{sec:conclusion}

The zero-coordinate moments of an MDS code separate the universal
interpolation counts from the moments that depend on prescribed
nonzero symbols. At strength two, only the minimum-weight covering
count remains undetermined, and its coefficient is nonzero at every
positive weight. This yields an equivalence between individual
$2$-design weights and an exact formula for their discrepancies.
At strength $t$, the same argument gives a discrepancy polynomial
of degree at most $t-2$ and determination by any $t-1$ design weights.

For DRS codes, the minimum-weight product criterion consequently
classifies each nonempty positive-weight shell separately. Deleting
the identity from the root-product domain gives a nonvanishing
faithful-character coefficient, which resolves the minimum-weight
$3$-design conjecture. Applying the classical shortening property
extends the obstruction to every weight below full support and
determines the exact strength there.

The full-support shell in the coprime case remains outside the
$3$-design classification proved here. The shortening index vanishes
at that boundary, so a further argument must use information retained
when no coordinate is zero. The character calculation above identifies
precisely where the present obstruction applies and where a different
count is needed.


\begin{thebibliography}{99}
\bibitem{AM}
E. F. Assmus, Jr. and H. F. Mattson, Jr.,
``New $5$-designs,''
\emph{Journal of Combinatorial Theory} \textbf{6} (1969), 122-151.

\bibitem{AGM}
E. F. Assmus, Jr., J.-M. Goethals, and H. F. Mattson, Jr.,
``Generalized $t$-designs and majority decoding of linear codes,''
\emph{Information and Control} \textbf{32}(1) (1976), 43-60.

\bibitem{Bonneau}
P. G. Bonneau,
``Weight distribution of translates of MDS codes,''
\emph{Combinatorica} \textbf{10} (1990), 103-105.

\bibitem{DMP}
A. A. Davydov, S. Marcugini, and F. Pambianco,
``On the weight distribution of the cosets of MDS codes,''
\emph{Advances in Mathematics of Communications}
\textbf{17}(5) (2023), 1115-1138.

\bibitem{DelsarteSchemes}
P. Delsarte,
\emph{An Algebraic Approach to the Association Schemes of Coding Theory},
Philips Research Reports Supplements, no.~10, 1973.
https://hdl.handle.net/2078.5/58922.

\bibitem{DelsarteFour}
P. Delsarte,
``Four fundamental parameters of a code and their combinatorial
significance,''
\emph{Information and Control} \textbf{23}(5) (1973), 407-438.

\bibitem{ElKhamyMcEliece}
M. El-Khamy and R. J. McEliece,
``The partition weight enumerator of MDS codes and its applications,''
preprint, 2005.
https://arxiv.org/abs/cs/0505054.

\bibitem{HP}
W. C. Huffman and V. Pless,
\emph{Fundamentals of Error-Correcting Codes},
Cambridge University Press, Cambridge, 2003.

\bibitem{Kosters}
M. Kosters,
``The subset sum problem for finite abelian groups,''
\emph{Journal of Combinatorial Theory, Series A} \textbf{120}(3)
(2013), 527-530.

\bibitem{KostersThesis}
M. Kosters,
\emph{Groups and Fields in Arithmetic},
Ph.D. thesis, Leiden University, 2014, Chapter 5, pp.~83-86
(reproduction of \cite{Kosters}). https://math.leidenuniv.nl/scripties/KostersPhD.pdf.

\bibitem{LiWan}
J. Li and D. Wan,
``Counting subset sums of finite abelian groups,''
\emph{Journal of Combinatorial Theory, Series A} \textbf{119}(1)
(2012), 170-182.

\bibitem{LN}
R. Lidl and H. Niederreiter,
\emph{Finite Fields}, 2nd ed.,
Encyclopedia of Mathematics and its Applications, vol.~20,
Cambridge University Press, Cambridge, 1997.

\bibitem{WZ26}
X. Wu and J. Zhou,
``A Complete Characterization of $q$-Ary $2$-Designs in
Doubly-Extended Reed--Solomon Codes'',
arXiv:2611.05861v1, November 4, 2026.

\bibitem{WZframework}
X. Wu and J. Zhou,
``Research on Linear Codes Holding $q$-Ary $t$-Designs'',
\emph{IEEE Transactions on Information Theory}, doi: 10.1109/TIT.2026.3742888, 2026.
arXiv:2603.12761v1, March 13, 2026.

\end{thebibliography}
\end{document}